\documentclass[aps, prx, floats, twocolumn, superscriptaddress, floatfix, longbibliography]{revtex4-2}

\setcitestyle{super}
\usepackage{multirow}
\usepackage{xcolor}
\usepackage{amsmath}
\usepackage{graphicx}
\usepackage[colorlinks=true, linkcolor = blue, urlcolor  = blue, citecolor = blue, anchorcolor = blue]{hyperref}

\usepackage{marginnote}

\definecolor{amber(sae/ece)}{rgb}{1.0, 0.49, 0.0}

\begin{document}

\title{\large Unmyelinated axons alter the biological interpretation of MRI biomarkers of demyelination}

\author{Ali Abdollahzadeh}
\affiliation{A.I. Virtanen Institute for Molecular Sciences, University of Eastern Finland, Kuopio, Finland}
\author{Ricardo Coronado-Leija}
\affiliation{Center for Biomedical Imaging, Department of Radiology, New York University School of Medicine, New York, NY, USA}
\author{Elizabeth Chasen}
\affiliation{Center for Biomedical Imaging, Department of Radiology, New York University School of Medicine, New York, NY, USA}
\author{Sungheon G. Kim}
\affiliation{Department of Radiology, Weill Cornell Medicine, New York, NY, USA}
\author{Dmitry S. Novikov}
\affiliation{Center for Biomedical Imaging, Department of Radiology, New York University School of Medicine, New York, NY, USA}
\author{Els Fieremans}
\thanks{Corresponding author:\\ els.fieremans@nyulangone.org}
\affiliation{Center for Biomedical Imaging, Department of Radiology, New York University School of Medicine, New York, NY, USA}

\begin{abstract} 
\noindent
Quantitative MRI methods are widely used to probe white matter microstructure; yet diffusion- and relaxation-based metrics are often interpreted primarily in terms of myelinated axons, implicitly assuming a negligible contribution from unmyelinated axons. 
Here, combining diffusion MRI (dMRI), relaxation-based MRI, and electron microscopy (EM) in the cuprizone mouse model of demyelination and recovery, we examine how unmyelinated axons shape MRI-derived microstructural indices, including axonal water fraction and $g$-ratio. 
To enable quantitative cellular-level comparisons, we developed automated tools for instance-level EM segmentation of white matter, allowing morphological analysis of myelinated and unmyelinated axonal populations. 
We find that, in the mouse model, diffusion-derived estimates of axonal water fraction reflect contributions from both axonal populations at sufficiently short diffusion times; accounting for unmyelinated axons removes an apparent bias and reveals that this metric corresponds to the total intra-axonal space rather than myelinated fibers alone. 
Similarly, MRI-derived $g$-ratio estimates correlate strongly with EM only when both axonal populations are included. 
We further demonstrate that within-voxel $g$-ratio heterogeneity introduces a systematic negative bias of up to $8\%$, increasing with the fraction of unmyelinated axons and variability in myelin thickness. 
Together, these findings change the biological interpretation of MRI biomarkers during demyelination by demonstrating that unmyelinated axons are a major determinant of MRI-derived white matter indices and that combining diffusion and relaxation contrasts enables a more accurate and interpretable quantification of demyelination.
They further highlight the need for quantitative biomarkers that account for axonal heterogeneity in both health and disease.
\end{abstract}

\maketitle

\section{Introduction}

\noindent Quantitative MRI is widely used to characterize white matter microstructure \textit{in vivo}, with particular relevance to demyelination and remyelination, where changes in myelin content and axonal integrity are central to disease progression and recovery~\cite{mackay_vivo_1994, fieremans_monte_2010, fieremans_white_2011, thiessen_quantitative_2013, filippi_magnetic_2014, weiskopf_advances_2015, mackay_magnetic_nodate, jelescu_vivo_2016, 
novikov_rotationally-invariant_2018, 
novikov_quantifying_2019,
papazoglou_insights_2024}. Measures derived from diffusion MRI (dMRI)~\cite{thiessen_quantitative_2013, jelescu_vivo_2016, novikov_quantifying_2019, papazoglou_insights_2024, sui_quantitative_2025}, transverse relaxation~\cite{mackay_vivo_1994, filippi_magnetic_2014, mackay_magnetic_nodate}, and magnetization transfer~\cite{schmierer_magnetization_2004, laule_magnetic_2007, weiskopf_advances_2015, sui_quantitative_2025} are commonly interpreted in terms of myelin loss, axonal density, or $g$-ratio, and have been extensively applied and validated in experimental models and human disease. These interpretations, however, implicitly rely on assumptions about the underlying axonal population that may not hold when myelin is disrupted.

Demyelinating insults, such as cuprizone exposure, cause oligodendrocyte death and myelin degeneration~\cite{matsushima_neurotoxicant_2001, praet_cellular_2014, schaffner_myelin_2023}, altering the myelination status of previously myelinated axons and increasing the proportion of ``effectively unmyelinated" axon segments in affected tracts. 
Whether this expansion reflects selective sparing of intrinsically unmyelinated axons~\cite{matsushima_neurotoxicant_2001, schaffner_myelin_2023} or axons with changed status~\cite{stidworthy_quantifying_2003, pfeiffer_structural_2019}, this shift in the statistical composition of axonal populations fundamentally changes the biophysical environment probed by MRI. Yet, most quantitative MRI models implicitly assume a homogeneous population of uniformly myelinated axons, overlooking this source of variability despite the extensive development of myelin-sensitive biomarkers~\cite{thiessen_quantitative_2013, filippi_magnetic_2014, weiskopf_advances_2015, stikov_vivo_2015, jelescu_vivo_2016, papazoglou_insights_2024}.

A particularly important implication of axonal heterogeneity in demyelinating tissue concerns the interpretation of the $g$-ratio. 
By definition, the $g$-ratio relates the inner and outer diameters of a myelinated axon~\cite{chomiak_what_2009, behanova_gacson_2022} and is therefore formally defined only for myelinated fibers, in both histological and MRI-based formulations~\cite{Novikov2012,stikov_vivo_2015, west_revised_2016, campbell_promise_2018}. 
In the presence of demyelination, where unmyelinated axons constitute an increasing fraction of the axonal population~\cite{matsushima_neurotoxicant_2001, schaffner_myelin_2023,stidworthy_quantifying_2003, pfeiffer_structural_2019}, voxel- or region-averaged $g$-ratio estimates no longer reflect per-axon myelin thickness alone but are influenced by changes in axonal composition. 
How this affects commonly used $g$-ratio measures during demyelination and remyelination is not sufficiently understood.

The cuprizone mouse model provides a highly reproducible system for studying demyelination and recovery~\cite{matsushima_neurotoxicant_2001, praet_cellular_2014} and testing the sensitivity and specificity of candidate biomarkers to myelin alterations. 
Cuprizone exposure causes selective oligodendrocyte loss and progressive demyelination, predominantly in the corpus callosum~\cite{skripuletz_-_2011}, followed by partial remyelination during continued intoxication and a second demyelination phase that emerges upon prolonged exposure.
Withdrawal of cuprizone after either acute or chronic treatment causes remyelination within several weeks~\cite{matsushima_neurotoxicant_2001, skripuletz_-_2011, zirngibl_oligodendrocyte_2022}.
We hypothesize that demyelination-induced changes in the relative abundance of myelinated and unmyelinated axons systematically alter the biological interpretation of quantitative MRI measures of axonal and myelin integrity. 

Validating quantitative MRI metrics requires direct comparison with histology~\cite{thiessen_quantitative_2013, jelescu_vivo_2016, salo_quantification_2018, molina_vivo_2020, fratini_multiscale_2020, papazoglou_insights_2024, abdollahzadeh_scattering_2025}. 
Sub-10-nanometer resolution electron microscopy (EM) enables resolving myelin and single-axon geometry in both myelinated and unmyelinated axons and quantifying their respective volume fractions~\cite{AbdollahzadehISMRM2023, abdollahzadeh_quantitative_2025}. 
However, model validation has largely focused on myelinated axons~\cite{abdollahzadeh_automated_2019, lee_along-axon_2019, behanova_gacson_2022, coronado-leija_volume_2024}, as automated instance segmentation of unmyelinated axons remains challenging due to their thin single-membrane boundaries and low contrast. 
Recent methodological advances have substantially expanded this landscape. 
dMRI models that explicitly describe axonal compartments~\cite{kaden_multi-compartment_2016, novikov_rotationally-invariant_2018, 
novikov_quantifying_2019,
veraart_noninvasive_2020, coelho_reproducibility_2022, veldmann_improving_2024} 
and orientation dispersion~\cite{tariq_binghamnoddi_2016, 
novikov_rotationally-invariant_2018, 
novikov_quantifying_2019,
coronado-leija_volume_2024}, together with improved estimation strategies~\cite{reisert_disentangling_2017, coelho_reproducibility_2022}
offer increased sensitivity to axonal composition. 
In parallel, recent developments in large-scale 3D EM acquisition~\cite{schlegel_whole-brain_2024} and automated instance segmentation~\cite{sheridan_local_2023, kirillov2023segment, schlegel_whole-brain_2024, archit_segment_2025} now enable a precise and comprehensive quantification of both axonal populations.

Here we reanalyze quantitative MRI and transmission EM (TEM) microstructural data acquired in a cuprizone model~\cite{jelescu_vivo_2016} with a distinct objective from previous work. 
Rather than validating MRI-derived biomarkers of demyelination~\cite{Novikov2012,filippi_magnetic_2014, weiskopf_advances_2015, stikov_vivo_2015, jelescu_vivo_2016, papazoglou_insights_2024}, we examine how demyelination-induced changes in axonal composition --- particularly shifts toward unmyelinated axons~\cite{matsushima_neurotoxicant_2001, schaffner_myelin_2023,stidworthy_quantifying_2003, pfeiffer_structural_2019} --- shape the interpretation of voxel-averaged MRI microstructure metrics, including $g$-ratio~\cite{stikov_vivo_2015, west_revised_2016, campbell_promise_2018}. 

We show that the dMRI-derived axonal water fraction $f$ of the Standard Model (SM) of diffusion in white matter~\cite{novikov_rotationally-invariant_2018,
novikov_quantifying_2019, coelho_reproducibility_2022} measured at short diffusion times, where unmyelinated axons remain effectively impermeable,
reflects the {\it total} intra-axonal volume rather than myelinated axons alone. 
Accounting for the combined contribution of myelinated and unmyelinated axons removes the apparent bias between $f$ and EM-derived axonal volume fractions~\cite{behanova_gacson_2022, AbdollahzadehISMRM2023}. On the other hand, the EM-derived myelin volume fraction strongly correlates with MRI measures sensitive to myelin content, including $R_2$ and magnetization transfer measures.
Combining these myelin-sensitive MRI measures with the diffusion-derived axonal water fraction $f$ yields an MRI-derived $g$-ratio that agrees with its EM-derived counterpart only when the contribution of unmyelinated axons is explicitly accounted for. Finally, we derive how intra-voxel $g$-ratio variability propagates through nonlinear averaging, leading to a systematic negative bias in effective $g$-ratio estimates. 
Together, these findings establish voxel-level heterogeneity, driven largely by unmyelinated axons, as a key determinant of MRI-derived measures of white matter integrity.

\section{Materials and methods} 

\noindent \textbf{Animal model and imaging}\\
Animal model, MR imaging, and EM preparation are detailed in Ref.~\cite{jelescu_vivo_2016} and are briefly summarized here.

\noindent \textbf{Animals}: An 18-week longitudinal study was conducted on 36 female C57BL/6 mice (8 weeks old at baseline; Taconic Biosciences, Hudson, NY), assigned equally to three diet groups of 12 animals. The control group (Ctl) received standard chow throughout the experiment. 
The Cpz6 and Cpz12 groups were fed chow supplemented with $0.2\%$ cuprizone for 6 and 12 weeks, respectively, followed by a 6-week recovery period on standard chow. MR imaging was performed \textit{in vivo} at baseline, week 6, week 12, and week 18. 
Following imaging, subsets of animals from each group were sacrificed at each time point (6, 12, and 18 weeks) for TEM imaging. Three mice from the Ctl group died during scanning, two at baseline and one at six weeks, so they were excluded.

All mice were treated in strict accordance with the National Institutes of Health Guide for the Care and Use of Laboratory Animals, and the experimental procedures were performed following the Institutional Animal Care and Use Committee at the New York University School of Medicine.\\ 

\noindent \textbf{MR imaging}: \textit{In vivo} MRI data of mice under anesthesia with isoflurane were acquired on a 7~T Bruker. MR imaging was restricted to a single 0.8-mm-thick mid-sagittal slice, which can be positioned reproducibly across animals and intersects callosal axons predominantly orthogonal to their main orientation. The slice was acquired at 112-$\mu$m in-plane resolution 

dMRI data were acquired along with $T_2$-weighted and magnetization transfer (MT) imaging. Diffusion-weighted images were acquired using a 2D GRASE (Gradient and Spin Echo) sequence~\cite{oshio_grase_1991, aggarwal_three-dimensional_2010} with $b$-values of 1 and 2 ms/$\mu$m$^2$ and 30 directions each, in addition to six $b = 0$ images ($T_R = 800\,$ms; $T_E = 49\,$ms; $\Delta = 16\,$ms; $\delta = 2\,$ms). $T_2$-weighted images were obtained using a multi-echo spin-echo sequence (32 echoes; $T_E$ spacing $= 7.1\,$ms; $T_R = 2\,$s; $N_A = 2$). 
Magnetization transfer ratio (MTR) maps were derived from two FLASH acquisitions ($T_R = 47\,$ms; $T_E = 5.5\,$ms; flip angle = 8$^\circ$; $N_A = 40$) with and without an MT saturation pulse (Gaussian shape; frequency offset = 1.5 kHz; duration = 12 ms; peak power = 4.7 $\mu T$).\\

\noindent \textbf{Electron microscopy}: Animals were anesthetized and perfused intracardially with heparinized phosphate-buffered saline, followed by fixation with $2\%$ formaldehyde and $2.5\%$ glutaraldehyde in phosphate buffer. Brains were extracted, and the splenium of the corpus callosum was dissected.
Tissue blocks were post-fixed in the same fixative, stained with osmium tetroxide, dehydrated, and embedded in resin.
Ultrathin sections were cut using an ultramicrotome and imaged with a Phillips CM12 TEM equipped with a Gatan digital camera, yielding electron micrographs of approximately $6 \times 9 \, \mu$m$^2$ at $2.36\times2.36$ nm$^2$ resolution.\\

\noindent \textbf{MR image processing}\\
To generate parametric maps, both the $T_2$-weighted images and the dMRI data were processed following the steps in the DESIGNER pipeline (\url{https://nyu-diffusionmri.github.io/DESIGNER-v2/})~\cite{chen_optimization_2024}.
Complex-valued images were denoised using MPPCA-Complex~\cite{Veraart2016, AdesAron2025}, corrected for Gibbs ringing~\cite{Kellner2015}, and registration~\cite{AVANTS2008} was used to correct geometric distortions induced by eddy currents as well as motion. 

$\boldsymbol{T_{2}}$ {\bf mapping} was performed to probe more specifically the $T_2$ of myelin and given the multi-compartmental nature of $T_2$ relaxation, we estimate $R_2=1/T_2$ as the slope at $T_E\to 0$
of the log-signal using a second-order cumulant expansion in $T_E$, $\ln S(T_E) \simeq \ln S_0 - R_2 T_E + c \cdot T_E^2$, using weighted linear least squares fitting.
To reduce the impact of stimulated echoes caused by imperfect refocusing pulses, we used only the even echoes from the 
$T_2$-weighted images, ranging from 14.168 ms to 226.688 ms in steps of 14.168 ms (excluding the last image due to low SNR). 

For \textbf{MTR mapping}, the MT$_{\rm on}$ image was registered to the MT$_{\rm off}$ image. 
MTR was then computed voxel-wise as: MTR = $({\rm MT}_{\rm off} - {\rm MT}_{\rm on}) \, / \, {\rm MT}_{\rm off} \times 100\%$, where MT$_{\rm off}$ is the signal without the MT pulse and MT$_{\rm on}$ is the signal with the pulse.
 
To generate the \textbf{SM maps} from dMRI, we first identified outliers~\citep{Collier2015} by the goodness-of-fit for diffusion kurtosis imaging (DKI) and replaced them with signals predicted from the DKI fit. The SM parameters are the axonal water fraction $f$, intra-axonal axial diffusivity $D_a$, extra-axonal parallel and perpendicular diffusivities $D_e^\parallel$ and $D_e^\perp$, and the free water compartment fraction $f_w$. The SM parameters were then estimated following Ref~\cite{coelho_reproducibility_2022}, with axial diffusivities $D_\mathrm{a}$ and $D_\mathrm{e}^{\parallel}$ constrained by an upper bound corresponding to the free diffusivity of water $3 \, \mu$m$^2$/ms at body temperature, while half this value was used to constrain the upper value of the radial extra-axonal diffusivity $D_\mathrm{e}^{\perp}$.  
Rotational invariants~\cite{novikov_rotationally-invariant_2018} up to $\ell = 2$ were used for SM parameter estimation, yielding the second-order rotational invariant $p_2$ of the fiber orientation dispersion, e.g., due to crossing fibers in a voxel.

The splenium of the corpus callosum was selected as the region of interest (ROI) as it exhibits robust and reproducible cuprizone-induced demyelination and rapid remyelination~\cite{stidworthy_quantifying_2003, steelman_demyelination_2012}, and encompasses the anatomical location sampled for 2D TEM. To define the ROI, we first manually drew a bounding box on the red channel (left-right orientation) of the color FA map to separate the splenium from the body of the corpus callosum.
Within this bounding box, the splenium was delineated automatically using a threshold of 0.8 times the maximum value, and the largest connected component of the resulting binary mask was retained. The corresponding mask was registered to $T_2$ and MTR spaces via the first $b = 0$ diffusion image and the $T_2$-weighted image with similar $T_E$ as the dMRI data, with registration verified visually. Mean and standard deviation of all MR-derived metrics were computed within the registered splenium ROI for each animal and time point.\\

\begin{figure}[t]
\centering
\includegraphics[width=1\linewidth]{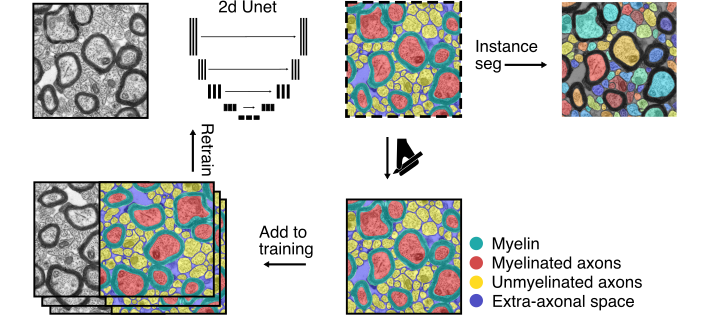}
\caption{\textbf{Segmentation of white matter microstructure.}
A standard 2D UNet was trained using a small set of manually labeled images until the network effectively memorized the training data. 
The trained model was then applied to unseen TEM images, and an expert corrected only the segmentation errors. 
The corrected image-label pairs were added to the training data, and the network was fine-tuned from its previous state. 
This iterative procedure was repeated, progressively expanding the training set while minimizing manual annotation. The final model was used for performing instance-level segmentation of white matter microstructure.}
\label{fig:pipeline}
\end{figure}

\noindent \textbf{2D EM microstructural segmentation}\\
We developed a DeepACSON-based~\cite{abdollahzadeh_deepacson_2021, abdollahzadeh_cylindrical_2021} segmentation pipeline to automatically segment high-resolution TEM images of white matter into myelin, myelinated axons, unmyelinated axons, and extra-axonal space, assigning a unique label to each compartment (\url{https://github.com/ComputationalNueroanatomy}).
The pipeline consists of three main steps, as shown in Fig.~\ref{fig:pipeline}: (i) incremental expansion of a training set; (ii) training a UNet~\cite{ronneberger_u-net_2015} for semantic segmentation of white matter microstructure; and (iii) their instance segmentation using a watershed-based method.

Training a UNet~\cite{ronneberger_u-net_2015} on a limited dataset poses challenges, yet modern architectures can fully ``memorize" even small datasets when trained with standard stochastic gradient descent~\cite{zhang_understanding_2021}. 
We exploited this property as a bootstrapping mechanism to generate a high-quality training set from minimal annotations. 
We began with a single manually labeled TEM image ($512\times512$ pixels), in which neurite boundaries and myelin were annotated. This image was augmented using randomized histogram distortions to ensure invariance to contrast and brightness gradients, as well as by applying affine transformations, random perspective warping, and horizontal/vertical flips. 

The initial UNet~\cite{ronneberger_u-net_2015} was trained to fully fit the augmented training set (Dice coefficient $\approx 1$). 
This overfitted network is not intended as a final predictor, but rather to generate initial segmentations for new, unlabeled images. 
Predictions were manually corrected only where necessary---typically in regions whose structural appearance deviated from the original training example---and the corrected images were added to the training set. This iterative refinement procedure progressively increased the diversity and coverage of the training data while minimizing the manual annotation effort. As the training dataset expanded, the network generalized to increasingly varied tissue appearances, requiring progressively less manual correction.

The model was trained using the cross-entropy loss with Softmax normalization. We set the batch size equal to the number of images in the training set. The training images were standardized to a mean equal to 0.456 and a standard deviation equal to 0.224. For the optimization, we used Adam optimizer~\cite{kingma_adam_2014} and set its initial learning rate $\alpha = 1 \times 10^{-4}$, the exponential decay rate for the first moment $\beta_1 = 0.9$, the exponential decay rate for the second-moment $\beta_2 = 0.999$, and the weight decay = $1 \times 10^{-5}$ (we use the same notation as in ref~\cite{kingma_adam_2014}). We utilized the trained UNet in its final state for the semantic segmentation of TEM images into their microstructural compartments. Finally, we applied a watershed segmentation for the instance segmentation of microstructural compartments.\\


\begin{figure}[t]
\centering
\includegraphics[width=1\linewidth]{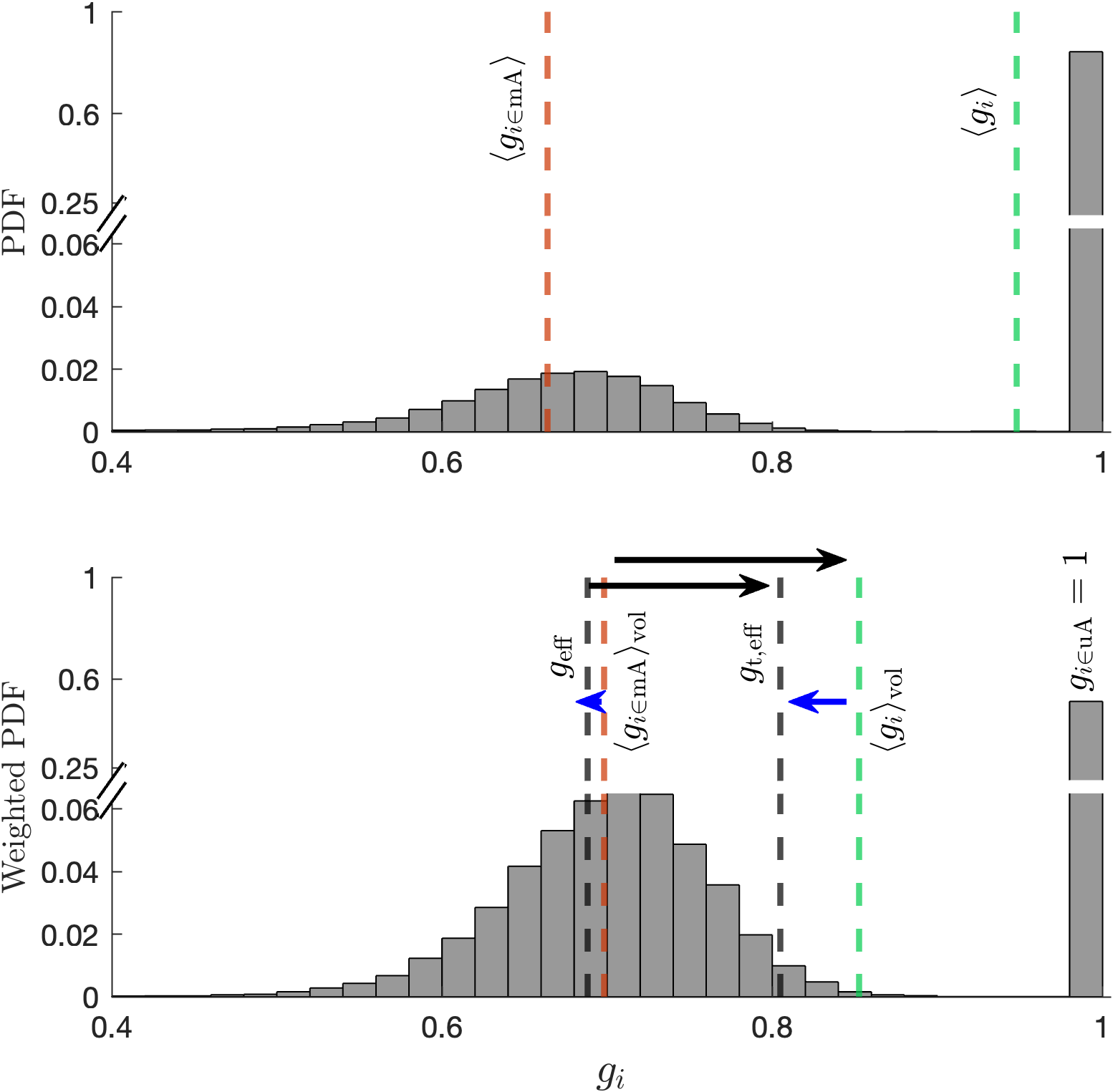}
\caption{\textbf{Histological and volume-weighted $g$-ratio distributions.}
The top panel shows the histological (unweighted) distribution of the $g$-ratio, while the bottom panel shows its volume-weighted MR-relevant analog. 
Distributions are obtained by pooling $g$-ratios from all axons of all histological sections and animals. 
The peak at $g_i = 1$ corresponds to formally including {\it unmyelinated axons} into the distributions, relevant for MR measurements (see text). 
Our main observation is that including the unmyelinated axons shifts the corresponding mean values notably to the right, Eq.~(\ref{eq:g_mA}) to Eq.~(\ref{eq:g_tA}), dashed black vertical lines (red to green), with the shift represented by the black arrows. 
Furthermore, the nonlinear averaging 
(\ref{nonlin_avg}) shifts the resulting effective $g$-ratio $g_{\rm eff}$ and $g_{\rm t, eff}$ a bit to the left relative to the respective distribution's mean $\langle g_i \rangle_{\rm vol}$, represented by the negative variance corrections in Eqs.~(\ref{eq:g_corr0}) and (\ref{eq:g_corr}). 
This negative correction (blue arrows) is smaller than the major increase due to including the unmyelinated axons (black arrows) in the calculation. 
The blue arrow correction is larger for the total effective $g$-ratio relative to that for myelinated axons alone, as the total distribution is broader.  
}
\label{fig:gratio_hist}
\end{figure}

\noindent \textbf{Statistical analysis}\\
For each histological measure, differences between diet groups were assessed separately at each time point using a linear mixed-effects model, with mouse as a random effect. For comparisons involving more than two groups, pairwise differences were tested using model-based contrasts, and adjusted $p$-values were computed using the Holm–Bonferroni correction. 
Effect sizes were estimated using Cohen’s d, noted as $d_\text{eff}$.

Correlations between EM- and MRI-derived metrics were assessed using weighted, nonparametric Spearman partial correlation analysis, accounting for mouse body weights, as cuprizone-treated mice exhibited reduced weight gain compared to controls. 
To characterize linear relationships between modalities, least-squares linear regression was performed between corresponding EM and MRI measures, and the resulting slope and intercept were used to assess proportionality and systematic bias.
All statistical tests were conducted at the two-sided $5\%$ significance level.

\section{g-Ratio definitions for heterogeneous axonal populations}

\noindent \textbf{EM-derived microstructural metrics}\\
The instance-level segmentation of white matter microstructure enables direct access to axonal diameters for individual myelinated and unmyelinated axons, denoted as $d_{\rm mA}$ and $d_{\rm uA}$. 
Our compartmental volumetric analysis includes the myelin volume fraction $\phi_{\rm m}$, the myelinated $\phi_{\rm {mA}}$ and unmyelinated axon volume fraction $\phi_{\rm {uA}}$, and the extra-axonal space volume fraction $\phi_{\rm {ex}}$, which are mutually exclusive and $\phi_{\rm m} + \phi_{\rm {mA}} + \phi_{\rm {uA}} + \phi_{\rm {ex}} = 1$. As these measurements were derived from 2D TEM images, area fractions were used as unbiased estimators of the corresponding volume fractions. 
We denote the total axonal volume fraction as $\phi_{\rm {tA}} = \phi_{\rm {mA}} + \phi_{\rm {uA}}$ (notations are summarized in Table~\ref{tbl:summary}).
When comparing EM-derived axonal volume fractions with the SM-derived axonal water fraction $f$, we normalized $\phi_{\rm mA}$, $\phi_{\rm uA}$, and $\phi_{\rm tA}$ by $1/(1-\phi_{\mathrm{m}})$ because $f$ reflects only ``MR visible" water and does not account for the myelin due to the fast $T_2$ relaxation of myelin water.
This normalization implicitly assumes comparable transverse relaxation properties across intra- and extra-axonal water pools. 
The differences in $T_2$ relaxation between intra- and extra-axonal compartments~\cite{veraart_te_2018, tax_measuring_2021} --- particularly under demyelinating conditions that we associated with the expansion of the extra-axonal space --- are not explicitly accounted for here and may cause slight deviations from linearity in the Cpz12-dMRI correlations. \\

\noindent \textbf{$g$-ratio: definitions and generalizations}\\
Originally, the $g$-ratio has been defined for individual myelinated axons only as the ratio $g=r/R$ of the inner to outer radii, implicitly assuming a perfectly circular axonal cross-section~\cite{rushton_theory_1951}. 
One can view it as a measure of the thickness $l_m = R-r=(1/g-1)r$ of the myelin sheath, consisting of many wrappings of a lipid bilayer. 

For non-circular axon cross-sections, both inner $r(\theta)$ and outer $R(\theta)$ radii would depend on the polar angle $0\leq \theta<2\pi$, even if the origin is taken to be at the geometric ``center of mass". 
To characterize myelin, one can either focus on the thickness $l_m(\theta)=R(\theta)-r(\theta)$ or on the ratio $g(\theta) = r(\theta)/R(\theta)$; it is rational to pick the metric that depends on $\theta$ the least. Here we assume that for the $i$-th axon, the $g$-ratio $g_i$ is approximately independent of $\theta$ (outer and inner boundaries are geometrically similar), thus the ratio between the inner and outer cross-sectional areas, $A_{{\rm in},i}$ and $A_{{\rm out},i}$, satisfies
\[
A_{{\rm in},i}= \int_0^{2\pi}\! \mathrm{d}\theta\, \tfrac12 r_i^2(\theta) =g_i^2 \, A_{{\rm out},i}
\]
independently of the cross-sectional shape.

In EM histology, where individual axons are resolved, we can measure this ratio $g_i$ at the single-axon level 
(top panel of Fig.~\ref{fig:gratio_hist}). 
We define the mean $g$-ratio of myelinated axons within each histological section as $\langle g_{i \in \mathrm{mA}}\rangle$; its average across sections for a given animal is $\left<g_i\right>$. 

From the MRI perspective, single-axon $g_i$ measurements are not accessible. 
Instead, an \emph{effective} $g$-ratio for myelinated axons $g_{\rm eff}$ is inferred\cite{stikov_vivo_2015} from voxel-averaged area (volume) fractions:  
\begin{align}
\label{eq:g_mA}
g_{\rm eff}
= \sqrt{\frac{\sum_{i\in{\rm mA}} A_{{\rm in},i}}{\sum_{i\in{\rm mA}} A_{{\rm out},i}}} =
\sqrt{\frac{\phi_{\rm mA}}{\phi_{\rm m}+\phi_{\rm mA}}} .
\end{align}
Eq.~(\ref{eq:g_mA}) nonlinearly pools the underlying intra-voxel distribution of $g_i$ into an effective 
\begin{equation} \label{nonlin_avg}
g_{\rm eff} = \frac1{\sqrt{\sum_i w_i g_i^{-2}}} 
\equiv \langle g_i^{-2}\rangle_{\rm vol}^{-1/2}
\ < \ \langle g_i \rangle_{\rm vol} \,,     
\end{equation}
where the weights 
\begin{equation} \label{weights}
w_i = {A_{{\rm in},i} \over \sum_j A_{{\rm in},j}} 
\end{equation}
(bottom panel of Fig.~\ref{fig:gratio_hist}).  
The largest axons dominate the volume-weighted average $\langle \dots \rangle_{\rm vol}$ due to the weights (\ref{weights}); a similar effect arises for MR-based axon diameters~\cite{burcaw_mesoscopic_2015, veraart_noninvasive_2020}.
Furthermore, the nonlinear averaging in Eq.~(\ref{nonlin_avg}) makes
the MR-based ratio (\ref{eq:g_mA}) {\it lower} than 
the volume-weighted mean $\langle g_i\rangle_{\rm vol}$. The latter can be seen by expanding in the deviations $\delta g_i = g_i - \langle g_i\rangle_{\rm vol}$ from the mean 
up to $(\delta g_i)^2$:  
\[
\big\langle \big(\langle g_i\rangle_{\rm vol}+\delta g_i\big)^{-2}\big\rangle_{\rm vol}^{\ -1/2} 
\approx \langle g_i\rangle_{\rm vol} \cdot \left< 1  + 3 \,{(\delta g_i)^2 \over \langle g_i\rangle_{\rm vol}^2} \right>_{\rm vol}^{-1/2}
\]
since $\langle \delta g_i \rangle_{\rm vol} \equiv 0$. 
As a result, the effective $g$-ratio  
\begin{align} \label{eq:g_corr0}
g_{\rm eff} \approx \langle g_i\rangle_{\rm vol} \cdot \left( 1 - \frac32 {\mathrm{var}_{\rm vol}\{g_i\} \over \langle g_i\rangle_{\rm vol}^2} \right) 
\end{align}
is negatively-biased relative to the volume-weighted mean. 

In demyelinating conditions, the axonal population is no longer composed solely of myelinated fibers. 
Unmyelinated or fully demyelinated axons can constitute a substantial fraction of the axonal population. 
We therefore define the total axonal volume fraction as $\phi_{\rm tA}=\phi_{\rm mA}+\phi_{\rm uA}$, and introduce a \emph{total effective} $g$-ratio
\begin{align} \label{eq:g_tA}
g_{\rm {t, eff}} = \sqrt{\frac{\phi_{\rm tA}}{\phi_{\rm {m}} + \phi_{\rm {tA}}}},
\end{align}
which includes unmyelinated axons with $g_{i\in\mathrm{uA}}=1$ by definition.
While $g_{\rm eff}$ reflects only the myelinated axon population, $g_{\rm {t, eff}} > g_{\rm eff}$ captures changes in the composition of the entire axon population during demyelination and remyelination.
Fig.~\ref{fig:gratio_hist} shows that the shift of $g_{\rm {t, eff}}$ relative to $g_{\rm eff}$ is very pronounced.   

Furthermore, the difference of the corresponding effective $g$-ratio from the volume-weighted distribution mean becomes more pronounced 
for $g_{\rm t, eff}$ under demyelinating conditions. Indeed, the presence of demyelinated and unmyelinated axons increases the variance of the intra-voxel $g$-ratio distribution through the additional contribution of axons with $g_i=1$. 
Hence, in the analog of Eq.~(\ref{eq:g_corr0}), 
\begin{align} \label{eq:g_corr}
g_{\rm t, eff} \approx \langle g_i\rangle_{\rm t,vol} \cdot \left( 1 - \frac32 {\mathrm{var}_{\rm t,vol}\{g_i\} \over \langle g_i\rangle_{\rm t,vol}^2} \right), 
\end{align}
the averages are now taken with respect to the {\it broader} distribution of $g_i$ for all axons, which includes a peak at $g_i=1$.  
In other words, the bias in effective $g$-ratio due to intra-voxel $g$-ratio variability becomes more pronounced in conditions involving myelin damage, where the variability increases due to the $g_i=1$ population (compare the two blue arrows in the right panel of Fig.~\ref{fig:gratio_hist}).

To summarize, including unmyelinated axons into the total $g$-ratio distribution makes the  MR-relevant $g$-ratio (\ref{eq:g_tA}) notably larger than that for the myelinated axons, Eq.~(\ref{eq:g_mA}). The second-order effect is that these quantities are a bit smaller than the mean values over the corresponding distributions due to the nonlinear character of the averaging (\ref{nonlin_avg}), with the negative corrections determined by the distribution variance (\ref{eq:g_corr0}) and (\ref{eq:g_corr}). The third-order effect is that this variance-based correction is more pronounced for the total distribution than for myelinated axons alone, as the total distribution is broader.

\begin{table}[t]
\centering
\caption{EM- and MR-derived microstructural metrics.}
\begin{tabular}{ll}
\hline
Metric & Definition \\
\hline
$\phi_{\rm m}$ & Myelin volume fraction\\
$\phi_{\rm mA}$ & Volume fraction of myelinated axons\\
$\phi_{\rm uA}$ & Volume fraction of unmyelinated axons\\
$\phi_{\rm ex}$ & Volume fraction of the extra-axonal space\\
$\phi_{\rm tA}$ & Total axonal volume fraction: $\phi_{\rm mA}+\phi_{\rm uA}$ \\
$d_{\rm mA}$ & Myelinated axon diameter\\
$d_{\rm uA}$ & Unmyelinated axon diameter\\
$g_{\rm eff}$ & Eq.~(\ref{eq:g_mA}) Effective $g$-ratio of myelinated axons \\
$g_{\rm t,eff}$ & Eq.~(\ref{eq:g_tA}) Effective $g$-ratio of total axons \\
$g_{\rm t,eff}^{*}$ & Eq.~(\ref{eq:g_corr}) Bias-corrected effective $g$-ratio \\
$\phi_{\rm m, MR}$ & MRI-derived myelin water fraction\\
$\phi_{\rm A, MR}$ & MRI-derived axonal water fraction\\
$g_{\rm MR}$ & MRI-derived $g$-ratio \\
$f$ & Standard Model axonal water fraction\\
\hline
\label{tbl:summary}
\end{tabular}
\end{table}


\begin{figure*}[t]
\centering
\includegraphics[width=0.85\linewidth]{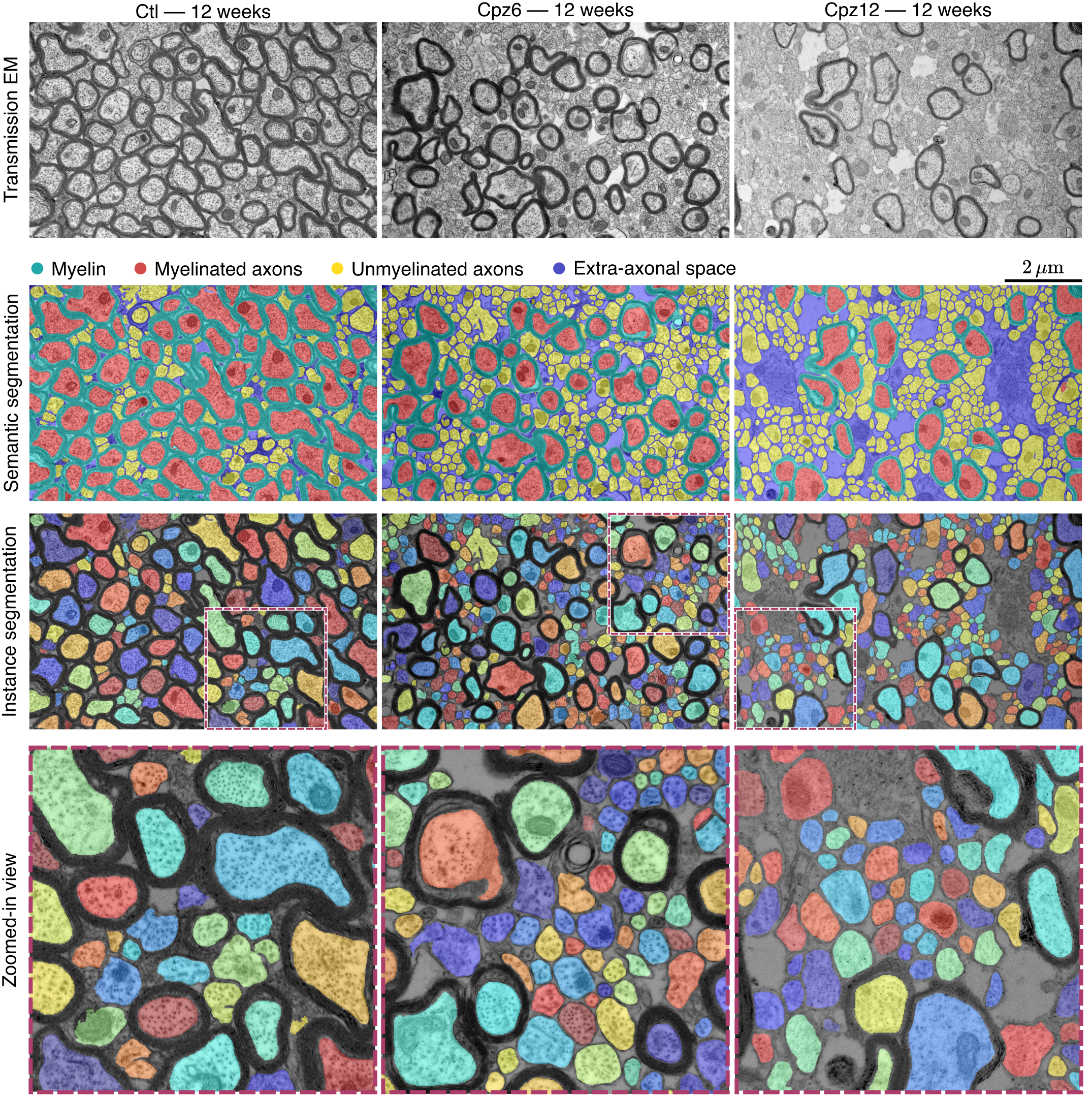}
\caption{\textbf{White matter microstructure segmentation in TEM images.} 
\textbf{(Top)} TEM images are acquired at ultra-high resolution with a voxel size of $2.36 \times 2.36$ nm$^2$: three representative TEM images at 12 weeks from the Ctl, Cpz6, and Cpz12 groups. 
\textbf{(Middle)} We applied a deep learning-based pipeline for the semantic segmentation of myelin, myelinated and unmyelinated axons, and the extra-axonal spaces; colors indicate microstructural classes. 
\textbf{(Bottom)} Instance segmentation and the zoomed-in view of myelinated and unmyelinated axons; colors are randomly assigned to individual axons.}
\label{fig:2dseg}
\end{figure*}

\begin{figure*}[t]
\centering
\includegraphics[width=0.95\linewidth]{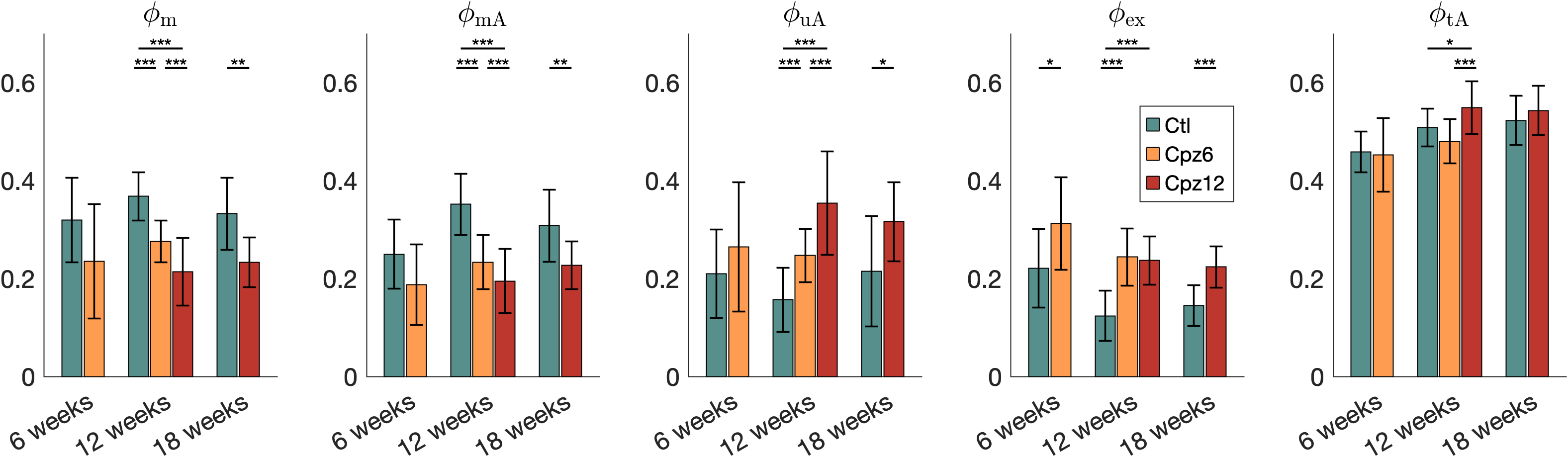}
\caption{\textbf{Morphological analysis of microstructural changes in cuprizone-induced demyelination.} 
Myelin- and axonal-related parameters derived from TEM images are shown for each group and time point.
Following cuprizone intoxication, demyelination is expected to reduce the myelin volume fraction $\phi_{\rm m}$ and the volume fraction of myelinated axons $\phi_{\rm mA}$, while increasing the volume fractions of unmyelinated axons $\phi_{\rm uA}$ and extra-axonal space $\phi_{\rm ex}$. The total axonal volume fraction $\phi_{\rm tA}$ is expected to remain approximately unchanged. Significant differences are asterisked: $*: p < 0.05; \, **: p < 0.01; \, ***: p < 0.001$.}
\label{fig:2dMorph}
\end{figure*}

\begin{figure}[t]
\centering
\includegraphics[width=0.9\linewidth]{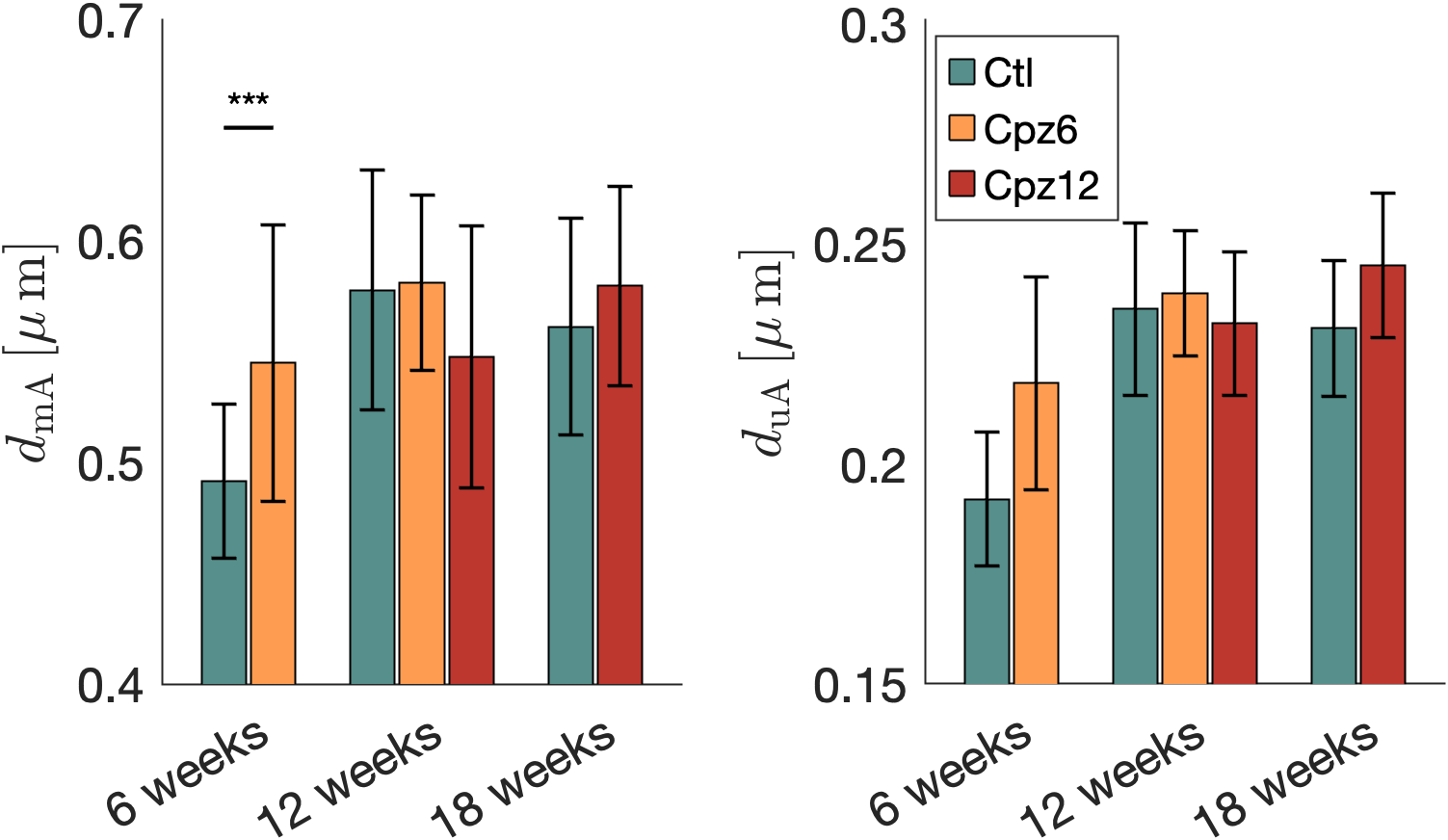}
\caption{\textbf{Axon diameter distributions are largely preserved across treatments.}
Distributions of axonal diameters for myelinated $d_{\rm mA}$ and unmyelinated $d_{\rm uA}$ axons show no differences between control and cuprizone-treated groups at most time points. 
A significant difference between Ctl and Cpz6 is observed at week 6; however, this effect is not sustained at later time points and is not accompanied by consistent shifts in diameter distributions. This relative consistency of axon diameter indicates that the observed changes in axonal volume fractions primarily reflect alterations in axonal population density and composition rather than pervasive diameter remodeling of individual axons.
Significant differences are asterisked: $*: p < 0.05; \, **: p < 0.01; \, ***: p < 0.001$.}
\label{fig:diam}
\end{figure}

\begin{figure*}[t]
\centering
\includegraphics[width=0.75\linewidth]{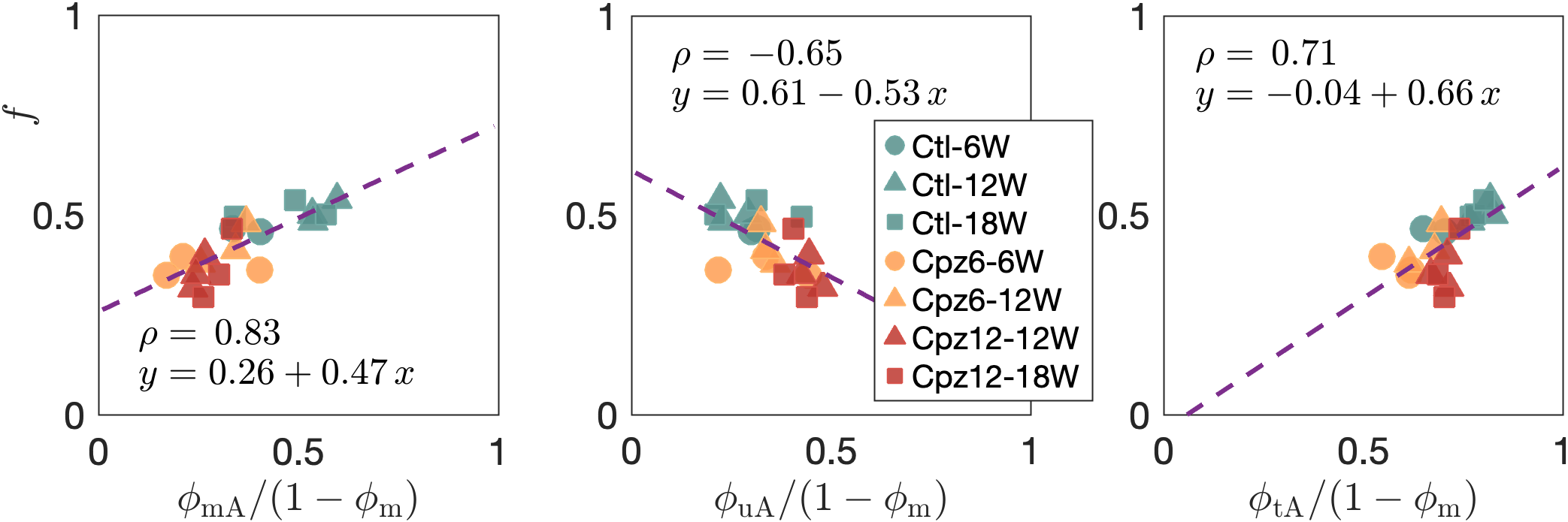}
\caption{\textbf{Correlation between the MR-derived intra-axonal water fraction $f$ and EM-derived microstructural volume fractions.} To enable direct comparison with MR-derived $f$, we normalize EM-derived volume fraction by $1/ (1 - \phi_{\rm {m}})$ as myelin does not contribute to the dMRI signal. 
Dashed lines indicate linear fits $(y = c_1 + c_2 x)$, where $x$ and $y$ are defined by the axis labels. The intra-axonal water fraction $f$ correlates strongly (Spearman correlation coefficient $\rho$) with the normalized volume fractions of myelinated axons $\phi_{\mathrm{mA}}$, unmyelinated axons $\phi_{\mathrm{uA}}$, and the total axonal population $\phi_{\mathrm{tA}}$. However, the linear fit between $f$ and $\phi_{\mathrm{tA}}$ exhibits a negligible intercept, in contrast to the plots showing $f$ vs $\phi_\mathrm{mA}$ and $f$ vs $\phi_\mathrm{uA}$, indicating that $f$ reflects contributions from both myelinated and unmyelinated axons rather than selectively capturing myelinated fibers alone.}
\label{fig:em_mr_corr}
\end{figure*}

\begin{figure*}[t]
\centering
\includegraphics[width=0.85\linewidth]{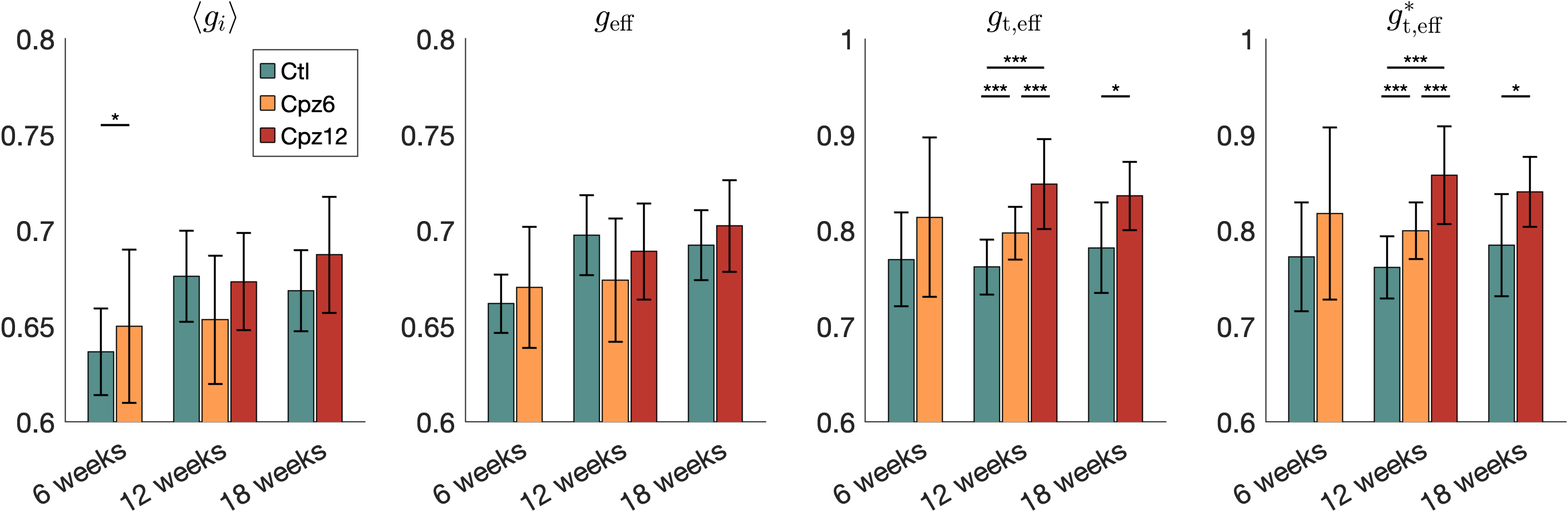}
\caption{\textbf{$g$-ratio analysis in cuprizone-induced demyelination.} 
Following cuprizone intoxication, the mean microscopic $g$-ratio $\langle g_i\rangle$ and the effective $g_{\rm eff}$ of myelinated axons Eq.~(\ref{eq:g_mA}) showed no significant differences among groups. 
In contrast, the effective $g_{\rm t,eff}$ of the total axonal population Eq.~(\ref{eq:g_tA}) and its bias-corrected $g_{\rm t,eff}^*$ analog Eq.~(\ref{eq:g_corr0}) (corrected for intra-voxel $g$-ratio variability) captured demyelination and increased significantly in cuprizone-treated groups at 12 and 18 weeks.
Significant differences are asterisked: $*: p < 0.05; \, **: p < 0.01; \, ***: p < 0.001$.}
\label{fig:gratio}
\end{figure*}

\begin{figure}[ht]
\centering
\includegraphics[width=1\linewidth]{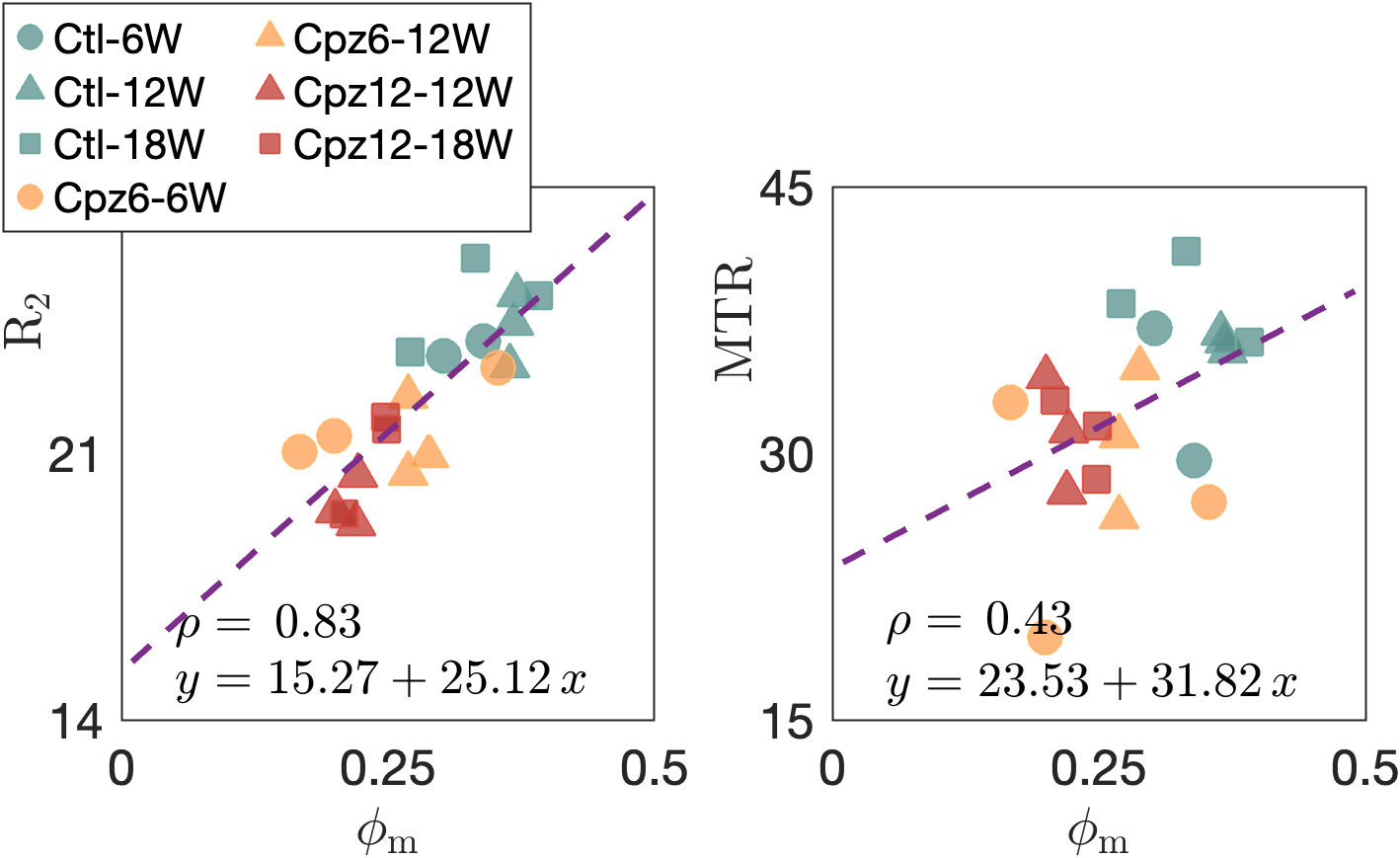}
\caption{\textbf{Correlation between MRI measures of myelination and histology.} 
$R_2$ shows a strong and significant correlation with the EM-derived myelin volume fraction $\phi_\mathrm{m}$, while MTR follows the same trend but with reduced sensitivity, resulting in only marginal statistical significance. 
Dashed lines indicate linear fits $(y = c_1 + c_2 x)$, where $x$ and $y$ are defined by the axis labels.}
\label{fig:em_mr_myelin}
\end{figure}

\begin{figure*}[ht]
\centering
\includegraphics[width=0.9\linewidth]{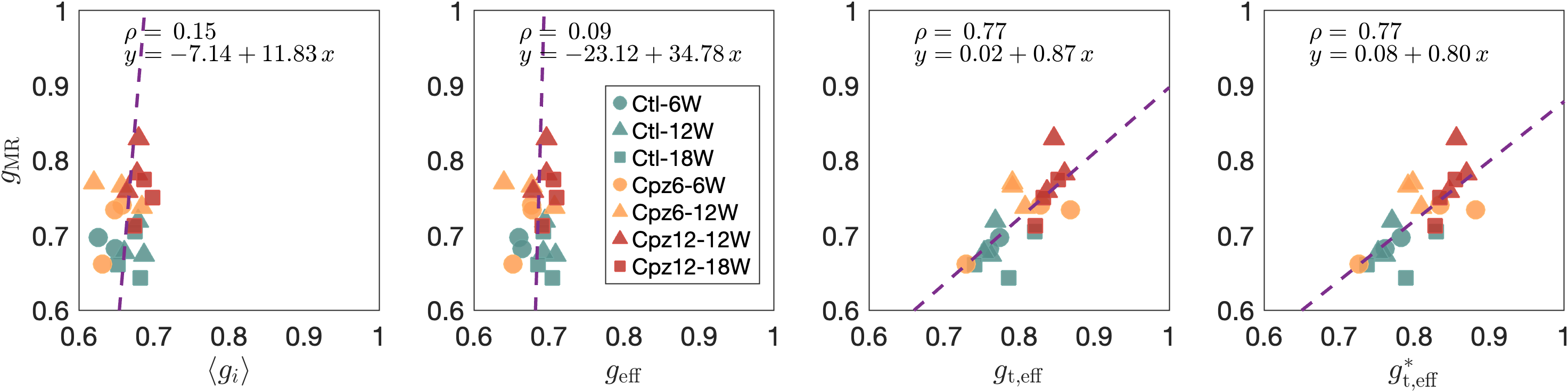}
\caption{\textbf{Correlation between histology- and MR-derived $g$-ratio.} The MR estimation of $g$-ratio correlates strongly with the histology measure of $g$-ratio when the contribution of unmyelinated axons is included, as in $g_\mathrm{tA}$ and the dispersion corrected $g_\mathrm{tA}^*$. Dashed lines indicate linear fits $(y = c_1 + c_2 x)$, where $x$ and $y$ are defined by the axis labels.}
\label{fig:em_mr_gratio}
\end{figure*}


\section{Results}

\noindent \textbf{Segmentation of white matter microstructure.}
Figure~\ref{fig:2dseg} shows the segmentation of TEM images of the splenium of the corpus callosum, acquired at an ultra-high resolution, 2.36 nm, enabling visualization of individual cellular and axonal membranes.
We performed semantic segmentation of the tissue into myelin, myelinated axons, unmyelinated axons, and extra-axonal space, using the trained UNet~\cite{ronneberger_u-net_2015} as shown in the middle row of Fig.~\ref{fig:2dseg}. 
Subsequently, each myelinated and unmyelinated axon has been assigned a unique label to achieve instance-level segmentation, enabling downstream morphometric and statistical analyses.\\

\noindent \textbf{Microstructural analysis of cuprizone exposure.}
Figure~\ref{fig:2dMorph} summarizes the quantitative analysis of TEM images and reveals microstructural changes consistent with the general expected trends in cuprizone-induced demyelination and recovery.

At 6 weeks of cuprizone exposure, we observed moderate reductions in $\phi_{\rm{m}}$ and $\phi_{\rm {mA}}$ ($d_{\rm {eff}} = -0.71$ and $-0.82$, respectively), accompanied by increases in $\phi_{\rm{uA}}$ and $\phi_{\rm {ex}}$ ($d_{\rm {eff}} = 0.39$ and $1.24$), with $\phi_{\rm tA}$ remaining largely unchanged ($d_{\rm eff} = -0.14$).

By 12 weeks, demyelination was markedly more pronounced. Both Cpz6 and Cpz12 groups exhibited significant decreases in $\phi_{\rm{m}}$ ($d_{\rm {eff}} = -8.73$ and $-14.20$) and
$\phi_{\rm {mA}}$ ($d_{\rm {eff}} = -2.80$ and $-7.60$). 
In contrast, we observed an increase in $\phi_{\rm{uA}}$ ($d_{\rm {eff}} = 3.50$ and $6.09$) and $\phi_{\rm {ex}}$ ($d_{\rm {eff}} = 3.53$ and $4.67$).
$\phi_{\rm tA}$ remained unchanged in the Cpz6 group ($d_{\rm eff} = -0.87$) and increased significantly only in the Cpz12 group ($d_{\rm eff} = 1.37$).

At 18 weeks, significant differences persisted between Cpz12 and control mice, with continued reductions in $\phi_{\rm{m}}$ and $\phi_{\rm {mA}}$ ($d_{\rm {eff}} = -1.70$ and $-1.59$), concomitant increases in $\phi_{\rm{uA}}$ and $\phi_{\rm {ex}}$ ($d_{\rm {eff}} = 1.15$ and $2.80$), and no change in $\phi_{\rm tA}$ ($d_{\rm {eff}} = 0.42$).
Notably, the opposite directions of change in $\phi_{\rm {mA}}$ and $\phi_{\rm {uA}}$ yield a total axon volume fraction $\phi_{\rm {tA}}$ approximately constant, suggesting overall axonal preservation despite extensive myelin loss. 

To evaluate the recovery, we examined within-group changes over time. Although Cpz6 mice remained significantly different from controls at 12 weeks and Cpz12 mice at 18 weeks, increases in $\phi_{\mathrm{m}}$ and $\phi_{\mathrm{mA}}$ from week 6 to week 12 in Cpz6 mice ($d_{\rm {eff}}^{\phi_{\mathrm{m}}} = 0.4$, $d_{\rm {eff}}^{\phi_{\mathrm{mA}}} = 0.64$), and from week 12 to week 18 in Cpz12 mice ($d_{\rm {eff}}^{\phi_{\mathrm{m}}} = 0.93$, $d_{\rm {eff}}^{\phi_{\mathrm{mA}}} = 1.54$), indicate partial remyelination following cuprizone withdrawal.\\

\noindent \textbf{The diameter of myelinated and unmyelinated axons.}
To assess whether the observed changes in axonal volume fractions were accompanied by alterations in axonal diameter, we compared the diameter distributions of myelinated and unmyelinated axons across groups and time points. Across all groups and time points, myelinated axons exhibited a mean diameter of $0.55 \pm 0.19~\mu$m, compared with $0.23 \pm 0.12~\mu$m for unmyelinated axons, indicating that myelinated axons are, on average, approximately twice as large. 
As shown in Fig.~\ref{fig:diam}, no significant differences in axonal diameter were observed for either population across treatment groups and time points, except for a transient difference in the diameter of myelinated axons between Ctl and Cpz6 mice at week 6. This effect was not sustained at later time points.\\

\noindent\textbf{Unmyelinated axons contribute to MRI-derived axonal water fraction.}
Here, we evaluate how intra-axonal water fraction $f$ correlates with the EM-measures of axonal volume fractions.
For comparison with $f$, EM measures were expressed relative to the MR-visible compartment by normalizing with $(1-\phi_{\rm m})$. 
We observed strong and statistically significant correlations ($\rho \ge |0.71|$, $p < 0.001$) between $f$ and all normalized EM-derived metrics (Fig.~\ref{fig:em_mr_corr}).
The negative correlation between $f$ and $\phi_{\mathrm{uA}}/(1-\phi_{\rm m})$ reflects their opposing changes during demyelination. 

Interestingly, the linear relationship between $f$ and $\phi_{\mathrm{tA}}/(1-\phi_{\rm m})$ exhibits a negligible intercept $(c_1 = -0.04)$, in contrast to the substantial non-zero intercept observed when $f$ is regressed against the $\phi_{\mathrm{mA}} /(1-\phi_{\rm m})$ $(c_1 = 0.20)$. 
This difference indicates that $f$ reflects contributions from both myelinated and unmyelinated axonal populations rather than selectively capturing myelinated fibers alone.
The slope deviates from unity, indicating a scaling effect between modalities.\\

\noindent \textbf{Effect of cuprizone exposure on $g$-ratio.}
Figure~\ref{fig:gratio} compares $g$-ratio definitions, derived from histology as described in \textit{Methods}, highlighting their sensitivity to demyelination and recovery.
The mean $\langle g_i \rangle$ and the effective $g_{\rm eff}$ in Eq.~\ref{eq:g_mA} reflect myelination within the myelinated axonal population.
Despite their different constructions, they fall within a similar numerical range and remain largely insensitive to demyelination and recovery, showing no significant changes across control and cuprizone groups at all time points.

In contrast, the total effective $g_{\rm t,eff}$ defined in Eq.~\ref{eq:g_tA} and its bias-corrected form $g_{\rm t,eff}^*$ in Eq.~\ref{eq:g_corr}, which corrects the negative bias from nonlinear averaging over heterogeneous $g$-ratios, explicitly incorporate the area-weighted contribution of unmyelinated axons, for which $g_{i \in \mathrm{uA}} = 1$. 
As a result, they attain a higher numerical range compared to $\left< g_i \right>$ and $g_{\rm eff}$.
Now, both $g_{\rm t,eff}$ and $g_{\rm t,eff}^{*}$ capture demyelination-related changes, showing significant increases in cuprizone-treated groups relative to controls at 12 weeks and remained higher in the Cpz12 group at 18 weeks.\\

\noindent\textbf{Unmyelinated axons contribute to experimental MR $g$-ratio.}
To estimate an experimental MR-derived $g$-ratio, denoted as $g_{\rm MR}$, we utilized $R_2$ and MTR that are, by design, most sensitive to myelin content. 
As shown in Fig.~\ref{fig:em_mr_myelin}, both metrics were evaluated against $\phi_{\mathrm{m}}$, where $R_2$ exhibited the strongest correlation ($\rho = 0.81, \, p < 0.001$), whereas MTR followed the same trend but with lower sensitivity ($\rho = 0.45, \, p = 0.05$). 

Based on the strong linear relationship between $R_2$ and $\phi_{\rm m}$ as shown in Fig.~\ref{fig:em_mr_myelin}, we estimated the MRI-derived myelin volume fraction as $\phi_{\rm m, MR} = (R_2 - 15.27) / 25.12$, obtained by inverting the fitted linear model. 
An MRI-derived axonal volume fraction was then estimated as $\phi_{\rm A, MR} = f \, (1-\phi_{\rm m, MR})$, where the factor $(1-\phi_{\rm m, MR})$ accounts for the non–MR-visible myelin compartment. 
The experimental MRI-derived $g_{\rm MR}$ was then calculated as 
\[
g_{\rm MR} = \sqrt{\frac{\phi_{\rm A, MR}}{\phi_{\rm m, MR} + \phi_{\rm A, MR}}}.
\] 
This construction yields an effective MRI-derived $g$-ratio analogous to Eqs.~\ref{eq:g_mA} and \ref{eq:g_tA}, whose correspondence to either the myelinated or total axonal population is evaluated below.

Comparison of $g_{\rm MR}$ with the $g$-ratio definitions (see Fig.~\ref{fig:em_mr_gratio}) introduced in \textit{Methods} showed poor correlation with both $\left<g_i\right>$ and $g_{\rm eff}$. 
In contrast, when unmyelinated axons were explicitly included, strong and significant correlations emerged between $g_{\rm MR}$ and both the total effective $g_{\rm t,eff}$ and $g_{\rm t,eff}^*$ $(\rho = 0.77,\, p < 0.001)$. 
This finding indicates that the MRI-derived $g$-ratio reflects the combined contribution of myelinated and unmyelinated axons rather than selectively reporting the myelin fraction of myelinated fibers alone.


\section{Discussion}

\noindent We studied white-matter microstructure in a cuprizone-induced model of demyelination and recovery, emphasizing the role of unmyelinated axons in shaping MR-derived microstructural metrics. 
The instance-level segmentation of myelinated and unmyelinated axons in high-resolution 2D TEM images, a task that is intrinsically difficult due to single-layer unmyelinated axons, allowed us to resolve axon-specific geometries and derive the full set of volume fractions, including that of myelin, myelinated and unmyelinated axons, and the extra-axonal space, together with $g$-ratio measures accounting for their intra-voxel variability.
Integration of these EM-based measurements with quantitative MRI revealed a nuanced picture of demyelination characterized by a pronounced loss of myelin and myelinated axons, accompanied by a relative expansion of unmyelinated axons and the extra-axonal compartment. 
Consistent with this structural reorganization, the SM-derived axonal water fraction  reflected contributions from both myelinated and unmyelinated axons, reinforcing its interpretation as a measure of {\it total} intra-axonal volume fraction rather than selectively probing myelinated fibers. This contribution of unmyelinated axons also substantially affected the MR-derived $g_{\rm MR}$-ratio.

The cuprizone-induced shift in axonal composition had a systematic impact on both histology- and MRI-derived measures of white matter integrity. 
In particular, this reorganization explains why the MRI-derived axonal water fraction correlated more strongly with the total axonal volume fraction than with the volume fraction of myelinated axons. dMRI is therefore sensitive to intra-axonal water from both myelinated and unmyelinated axons and should not be interpreted as a measure specific to myelinated axons alone. 

Importantly, this interpretation applies to the relatively short diffusion time used in the present study ($\Delta$ = 16 ms). At diffusion times shorter than the exchange time, water associated with unmyelinated axons contributes directly to the MR-measured axonal water fraction. As diffusion time exceeds the exchange time, inter-compartmental exchange between unmyelinated axons and extra-axonal space progressively transfers this contribution to the extra-axonal compartment~\cite{Coronado-LeijaISMRM2025, Coronado-LeijaISMRM2026}. Therefore, the contribution of unmyelinated axons to the measured axonal water fraction decreases with increasing diffusion time. This suggests that clinical measurements, which typically operate at longer diffusion times, may be less sensitive to the distinct contribution of unmyelinated axons than shorter-time preclinical experiments. Distinguishing between myelinated and unmyelinated axons is critical in pathological contexts in which myelin loss occurs without overt axonal degeneration, such as the early demyelinating or remyelinating stages of multiple sclerosis~\cite{lassmann_multiple_2018, schaffner_myelin_2023}, where apparent reductions in myelin content may coexist with preserved total axonal density.

The EM-derived metrics exhibited the expected trends of cuprizone intoxication and recovery. 
Following the chronic cuprizone exposure, the volume fraction of myelin and myelinated axons decreased, while the volume fraction of unmyelinated axons increased. 
During the recovery phase, these effects were partially reversed, consistent with prior reports of demyelination and remyelination dynamics in the cuprizone model~\cite{matsushima_neurotoxicant_2001, stidworthy_quantifying_2003, lindner_sequential_2008, xie_rostrocaudal_2010, steelman_demyelination_2012} and the increase of the population of intrinsically unmyelinated axons as an adaption mechanism~\cite{pfeiffer_structural_2019} or, at minimum, a selective sparing of unmyelinated axons from cuprizone toxicity~\cite{matsushima_neurotoxicant_2001}.
The near-constancy of the total axonal volume fraction across treatment conditions suggests that cuprizone primarily affects the myelin sheath rather than causing gross axonal loss. 

The analysis of axonal diameters further supports this interpretation (for a 3D EM study that quantifies myelinated and unmyelinated axons along axonal trajectories, see Ref.~\cite{AbdollahzadehISMRM2023}). The absence of systematic changes in axonal diameters indicates that the observed alterations in axonal volume fractions are not driven by swelling, shrinkage, or caliber remodeling of individual axons. Instead, they primarily reflect changes in axonal population density and composition within the tissue. Note that we do not distinguish between axons that were intrinsically unmyelinated and those that became unmyelinated following demyelination. 

The myelin-sensitive MR metrics $R_2$ and MTR were sensitive to cuprizone-induced demyelination and recovery at all time points, in line with previous studies~\cite{song_demyelination_2005, zaaraoui_monitoring_2008, wu_mri_2008, thiessen_quantitative_2013, falangola_histological_2014}, with $R_2$ showing the strongest correlation. However, MTR showed only a modest correlation with $\phi_\mathrm{m}$ in our data. Although previous studies reported significant correlations between MTR and either myelin basic protein content~\cite{zaaraoui_monitoring_2008} or myelin volume fraction~\cite{thiessen_quantitative_2013}, MTR is not specific to myelin. In particular, significant correlations between MTR and the volume fraction of non-myelinated cells have also been reported~\cite{thiessen_quantitative_2013}, yet it can be sensitive to inflammation and edema~\cite{gareau_magnetization_2000, vavasour_is_2011}. These limitations in myelin specificity should therefore also be considered when interpreting $g_{\rm MR}$-ratio, which combines myelin- and axonal water fraction.

Despite pronounced reductions in myelin volume fraction, the mean $g$-ratio $\left< g_i \right>$ and the effective $g_{\rm eff}$ of myelinated axons in Eq.~\ref{eq:g_mA} remained largely unchanged across treatment groups, underscoring the maintained ratio between the inner and outer diameters in \textit{surviving} myelinated axons. 
In contrast, the total effective $g_{\rm t,eff}$ in Eq.~\ref{eq:g_tA}, which explicitly incorporates the area-weighted contribution of unmyelinated axons ($g_{i\in \rm uA} = 1$), increased significantly during demyelination and corresponds best to the experimental MR-derived $g_{\rm MR}$ as measured from quantitative dMRI and $T_2$-MRI. This increase reflects the growing fraction of unmyelinated axons and links the total effective $g$-ratio to the effective redistribution of myelination within an MRI voxel, rather than to changes in per-axon myelin thickness.

\begin{figure}[t]
\centering
\includegraphics[width=0.9\linewidth]{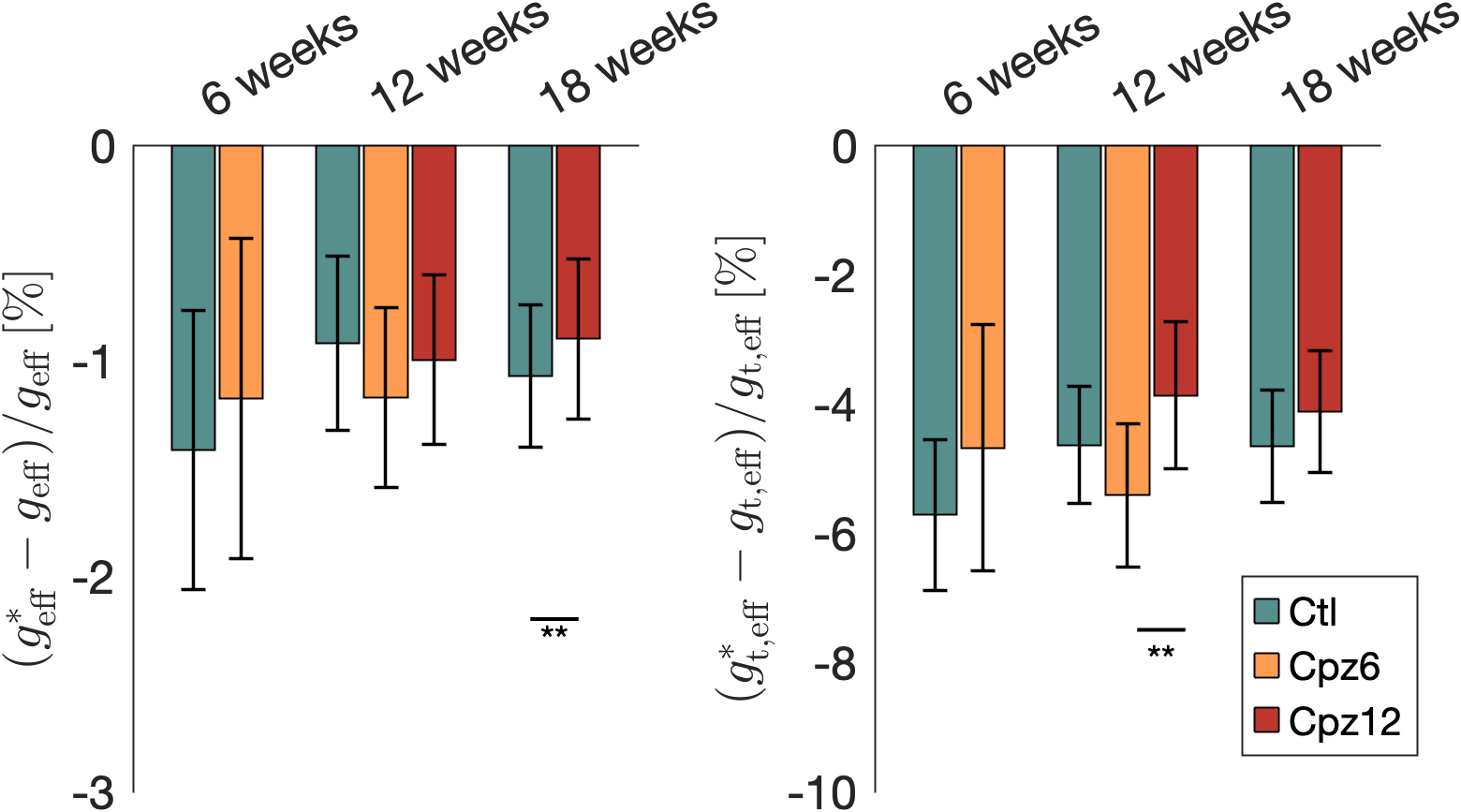}
\caption{\textbf{Within-voxel variability in the $g$-ratio results in a systematic negative bias in the MR-derived g-ratio.}
Relative error in effective $g$-ratio estimates arising from discarding within-voxel $g$-ratio variability, shown separately for myelinated axons $g_{\mathrm{MR}}$ and the total axonal population $g_{\mathrm{t,MR}}$. 
The bias is larger for $g_{\mathrm{t,MR}}$, reflecting the increased variance introduced by including unmyelinated axons with $g_i = 1$. 
In some comparisons, this bias reaches statistical significance.
Significant differences are asterisked: $*: p < 0.05; \, **: p < 0.01; \, ***: p < 0.001$.}
\label{fig:gratio_err}
\end{figure}

The effective $g_{\rm eff}$ combines estimates of myelin and axonal volume fractions to yield a voxel-wise index of myelination~\cite{stikov_vivo_2015, campbell_promise_2018}. 
Yet, this model implicitly assumes a uniform $g$-ratio within each voxel, an approximation that neglects intra-voxel heterogeneity in myelin thickness and the presence of unmyelinated axons. 
Such heterogeneity biases effective $g$-ratio estimates toward lower values~\cite{campbell_promise_2018}. 
As shown in Fig.~\ref{fig:gratio_err}, this bias becomes substantial in tissue with a high fraction of unmyelinated or demyelinated axons, reaching up to $8\%$ and potentially introducing apparent group differences or obscuring true biological effects. 

We derived an analytical correction that links the variance of the $g$-ratio distribution to the macroscopic bias observed in both $g_{\rm eff}$ and $g_{\rm t,eff}$. This framework establishes a quantitative connection between microscopic $g$-ratio variability and its measurable effective counterpart. In practice, however, applying this correction led to overcompensation: the bias-corrected measure $g_{\rm t,eff}^{*}$ increased the mismatch with $g_{\rm MR}$ (bias $0.02 \to 0.08$). This overshoot suggests that the magnitude of $\mathrm{var}_{\rm vol}\{g_i\}$ required to reconcile MRI- and EM-derived $g$-ratio estimates is larger than what is observed in the data, implying that $g$-ratio heterogeneity alone cannot fully explain the observed discrepancy between modalities.

In conclusion, the diffusion-derived axonal water fraction reflects the total axonal volume fraction at short diffusion times, encompassing both myelinated and unmyelinated axons, as confirmed by its strong correspondence with EM-derived microstructural measures. The same principle applies to the $g$-ratio: accurate interpretation requires inclusion of unmyelinated axons, whose contribution substantially influences voxel-averaged estimates. Moreover, intra-voxel heterogeneity in $g$-ratio distributions systematically underestimates the effective $g$-ratio, with the magnitude of this reduction increasing alongside the fraction of unmyelinated axons and the $g$-ratio variability. 
Together, these findings demonstrate that unmyelinated axons are integral to the microstructural contrast detected by MRI and must be considered explicitly when quantifying axonal and myelin integrity in both healthy and demyelinated white matter.\\

\section*{Data availability}
The high-resolution EM data are publicly available through the Open Science Framework at \url{https://osf.io/adrcs/overview}. 
DeepACSON is publicly available at \url{https://github.com/ComputationalNueroanatomy}, and the DESIGNER pipeline is available at \url{https://nyu-diffusionmri.github.io/DESIGNER-v2}.

\section*{Acknowledgments}
We thank Prof. James L. Salzer and Prof. Jiangyang Zhang for their discussions and feedback on this work. We thank Hong-Hsi Lee for his feedback. We also thank Ileana O. Jelescu and Magdalena Zurek for the data used in this study.

\section*{Funding}
The research was supported by the NIH under awards R01 NS088040 and R21 NS081230, and by the Irma T. Hirschl fund. This research was supported by the Research Council of Finland (grant $\#$360360) awarded to A.A. The research was performed at the Center of Advanced Imaging Innovation and Research (CAI2R, www.cai2r.net), a Biomedical Technology Resource Center supported by NIBIB with the award P41 EB017183.

\section*{Author contributions}
A.A., D.S.N., and E.F. conceived the project and designed the study. 
A.A. conducted EM image analysis and analyzed the data. 
R.C.-L. conducted MRI processing and analysis. 
E.C. contributed to EM analysis.
S.G.K contributed to data acquisition.
A.A. and D.S.N. derived the analytical expressions.
A.A. and E.F. wrote the manuscript. 
D.S.N. and E.F. supervised the project. 
All authors commented on and approved the final manuscript.

\section*{Competing interests}
The authors declare no competing financial interests.

\bibliography{ref}

\end{document}